\documentclass[preprint,12pt]{elsarticle}

\usepackage{amssymb}
\usepackage{amsmath}
\usepackage[utf8]{inputenc}
\usepackage[T1]{fontenc}
\usepackage{graphicx}
\usepackage{array}
\usepackage{hyperref}
\usepackage{xcolor}
\usepackage{booktabs}

\usepackage[tableposition=top]{caption}

\newcommand{\hi}[1]{#1} 

\journal{Computers in Biology and Medicine}

\begin{document}

\begin{frontmatter}

\title{Automated Identification of Cell Populations in Flow Cytometry Data with Transformers}

\author[1,2]{Matthias Wödlinger}
\ead{mwoedlinger@cvl.tuwien.ac.at}

\author[1,2]{Michael Reiter}
\author[1]{Lisa Weijler}
\author[2]{Margarita Maurer-Granofszky}
\author[2]{Angela Schumich}
\author[3]{Elisa O. Sajaroff}
\author[4]{Stefanie Groeneveld-Krentz}
\author[3]{Jorge G.Rossi}
\author[4]{Leonid Karawajew}
\author[5]{Richard Ratei}
\author[2]{Michael N. Dworzak}

\address[1]{TU Wien, Vienna, Austria}
\address[2]{St Anna Children's Cancer Research Institute, Vienna, Austria}
\address[3]{Cellular Immunology Laboratory, Hospital de Pediatria “Dr. Juan P. Garrahan”, Buenos Aires, Argentina}
\address[4]{Department of Pediatric Oncology/Hematology, Charité Universitätsmedizin Berlin, Berlin, Germany}
\address[5]{Department of Hematology, Oncology and Tumor Immunology, HELIOS Klinikum Berlin-Buch, Berlin, Germany}

\begin{abstract}
Acute Lymphoblastic Leukemia (ALL) is the most frequent hematologic malignancy in children and adolescents. A strong prognostic factor in ALL is given by the Minimal Residual Disease (MRD), which is a measure for the number of leukemic cells persistent in a patient. Manual MRD assessment from Multiparameter Flow Cytometry (FCM) data after treatment is time-consuming and subjective. In this work, we present an automated method to compute the MRD value directly from FCM data. We present a novel neural network approach based on the transformer architecture that learns to directly identify blast cells in a sample. We train our method in a supervised manner and evaluate it on publicly available ALL FCM data from three different clinical centers. \hi{Our method reaches a median $F_1$ score of $\approx 0.94$ when evaluated on $519$ B-ALL samples and shows better results than existing methods on $4$ different datasets}.\footnote{Our code is available on Github: \url{https://github.com/mwoedlinger/flowformer}}
\end{abstract}



\begin{keyword}
multiparameter flow cytometry  \sep automated gating \sep deep learning \sep self-attention.
\end{keyword}

\end{frontmatter}


\section{Introduction}
\label{sec: introduction}
Acute Lymphoblastic Leukaemia (ALL) is a malignant disorder of lymphoid progenitor cells. It is the most frequent hematologic malignancy in children and adolescents and treated patients show relapse rates of $15-20\%$ \cite{pui2008acute}. A means of tracking the progress of treatment is provided by the Minimal Residual Disease (MRD), which is the fraction of remaining leukemic cells ({\it blast} cells) after therapy. Low MRD values in early stages of treatment have been shown to be powerful predictors for better outcomes \cite{campana2010minimal}. For this reason, the correct assessment of MRD values is an important part of modern treatment methods.  
\begin{figure*}[t]
	\centering
	\includegraphics[width=\textwidth]{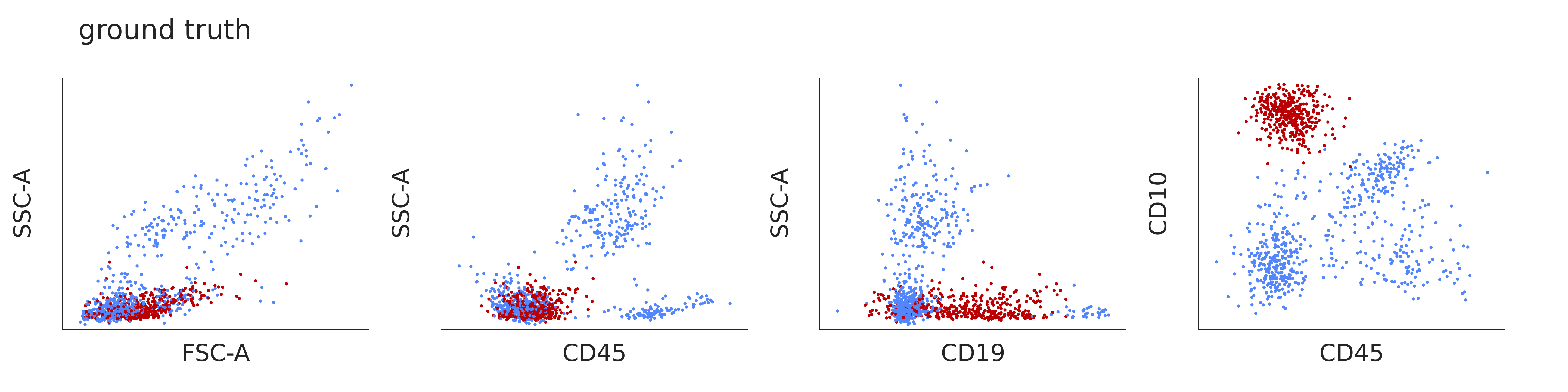}\\
	\includegraphics[width=\textwidth]{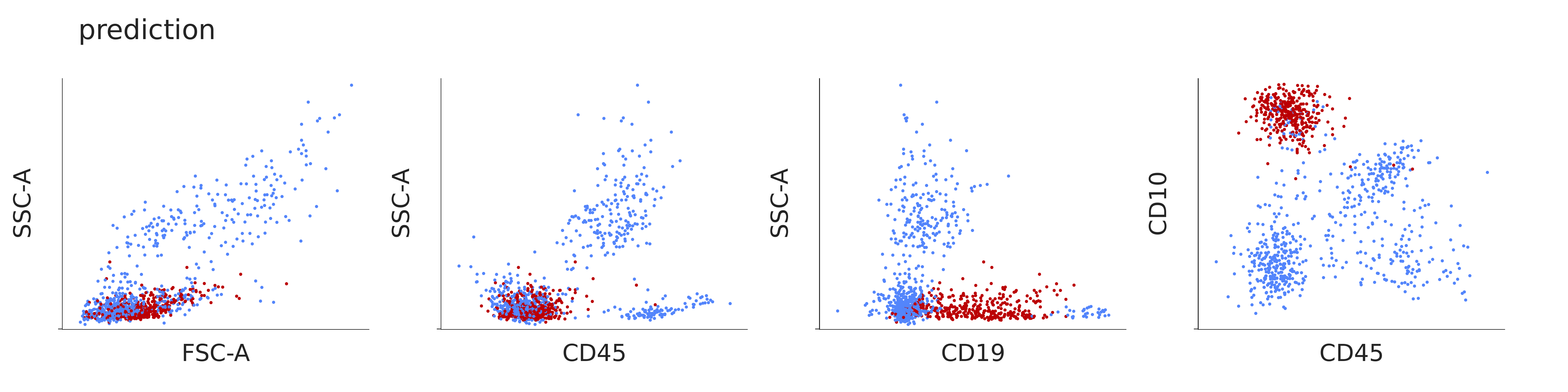}
	\caption{\hi{An example prediction (bottom row) of our method and the corresponding manual labeling (top row). Red dots denote blast and blue dots non-blast cells. Every plot shows a different $2$-dimensional projection of the same underlying FCM data sample. For this visualisation we randomly sampled $5000$ cells from a sample from the {\it bue} dataset. The prediction is from a model trained on {\it vie14} (see Tab. \ref{tab: data} for a description of the datasets).}}
	\label{fig: example}
\end{figure*}
\paragraph{Flow Cytometry}
\label{par: FCM}
Multiparameter Flow Cytometry (FCM) provides a reliable way to obtain MRD values during treatment \cite{dworzak2002prognostic}. In this process, a blood or bone marrow sample of a patient is stained with a specific combination of fluorescence-labelled antibodies that bind to their respective cell antigens. In the Flow cytometer machine, the cells are then illuminated by a selection of lasers that allow the detection and measurement of physical properties (granularity, size) as well as biological properties through detection of the antibodies if attached to respective antigens. The resulting data for a single cell (an {\it event}), is a collection of measurements of cell surface marker concentrations (see Figure \ref{fig: example} for an example of FCM data as seen during clinical routine). However, manual analysis of FCM data is time-consuming, subjective and dependent on the operator's experience.
\paragraph{Contribution}
To tackle the shortcomings of manual gating several methods have been proposed that allow automated FCM analysis. The structure of FCM data samples however proves to be challenging for neural network-based approaches as these often require data points on a grid (e.g. convolutional neural networks for a 2d grid or recurrent neural networks for sequences). Some methods \cite{scheithe2019monitoring,licandro2018wgan} circumvent this problem by applying neural networks on single cells instead of samples, however, these approaches can only learn static decision boundaries and are not able to capture global sample information. In this work, we present a novel method for the detection of blast cells and MRD quantification that is capable of capturing long-range information in the full data space by attending to all events in a sample at once. Our method consists of a neural network based on the transformer \cite{vaswani2017attention,lee2019set} architecture, that learns gating directly from FCM data. This allows fast inference and easy adaption to new data. The remaining paper is structured as follows: After a discussion of the related work in the next section \ref{sec: related work}, we present our approach in section \ref{sec: methods} and show the results in section \ref{sec: experiments}.
%
\section{Related Work}
\label{sec: related work}
In manual gating methods, cells are identified based on $2$-dimensional projections of the (higher-dimensional) FCM data. Automated methods, on the other hand, can utilize the full parameter space. Typically, these methods aim to assign the correct population to every single cell. This produces an output similar to manual gating. This output can then be used directly in clinical routine (for example for MRD quantification) or as a starting point for further data analysis. We present a selection of related methods for automated FCM analysis and then discuss recent progress in Transformer \cite{vaswani2017attention} related neural networks. In particular, the progress with respect to complexity and memory footprint reduction for long sequences is discussed. Being able to process longer sequences is essential for FCM data where samples typically contain $10^5$ to $10^6$ cells.
\paragraph{Automated FCM analysis}
Several works formulate automated FCM analysis as an unsupervised learning problem by adopting non-parametric density estimation or clustering methods \cite{sorensen2015immunoclust,aghaeepour2013critical}. A line of research that recently showed good results in both the unsupervised \cite{naim2014swift,dundar2014non,johnsson2016bayesflow} and supervised \cite{reiter2016clustering,reiter2019automated} setting are Gaussian Mixture Models (GMM). In SWIFT \cite{naim2014swift} the conventional GMM algorithm is adapted to better detect rare sub-populations; BayesFlow \cite{johnsson2016bayesflow} employs a hierarchical Bayesian model were expert knowledge can be incorporated through informative priors. \cite{reiter2016clustering} accounts for inter-sample variation with a supervised approach where GMMs are matched to GMMs of a labelled reference dataset. The method is advanced in \cite{reiter2019automated} where a closed form optimization in the fitting process is introduced. \hi{Deep learning has been successfully applied to automated processing of image cell data \cite{iqbal2021deep, iqbal2021deep}, however, apart from imaging FCM applications \cite{nissim2020real,eulenberg2017reconstructing, li2018accurate}, few examples of successful application of deep neural networks to FCM data exist}. In  \cite{licandro2018wgan,scheithe2019monitoring,li2017gating} neural networks based on fully connected layers are presented that work on single events. These methods can only learn fixed decision boundaries to separate biologically meaningful sub-populations. Only recently in \cite{zhao2020hematologist} a method has been proposed to circumvent this problem by transforming FCM data to image space and processing it with a learned CNN.
\paragraph{Transformers}
The original Transformer paper \cite{vaswani2017attention} introduced a neural network layer that allows capturing of long-range information. In theory, the layer is capable of capturing global information, however, due the complexity of both memory and time being quadratic in the input length $\mathcal{O}(N^2)$ the authors restrict input sequences to $2048$ tokens. One solution to this is provided by the Reformer \cite{kitaev2020reformer}. Here the authors use locality-sensitive hashing to restrict the attention to nearby positions which reduces the time complexity to $\mathcal{O}(N \log N)$. While this results in a similar performance to the original transformer for many tasks, it restricts the attention to the local neighbourhood. Another line of research is given by models\cite{choromanski2020rethinking,katharopoulos2020transformers} that aim to achieve linear complexity $\mathcal{O}(N)$ by approximating the Softmax function in the self-attention layer with a kernel which allows factorizing the computation of the attention matrix. Most related to our application are Set Transformers \cite{lee2019set}, a type of transformer architecture specifically designed for set inputs where the order of inputs is not relevant. These networks achieve linear complexity with the sequence length $\mathcal{O}(N)$ by applying the idea of inducing points from the theory of Gaussian processes.
%
\section{Methods}
\label{sec: methods}
\begin{figure}
    \includegraphics[width=\textwidth]{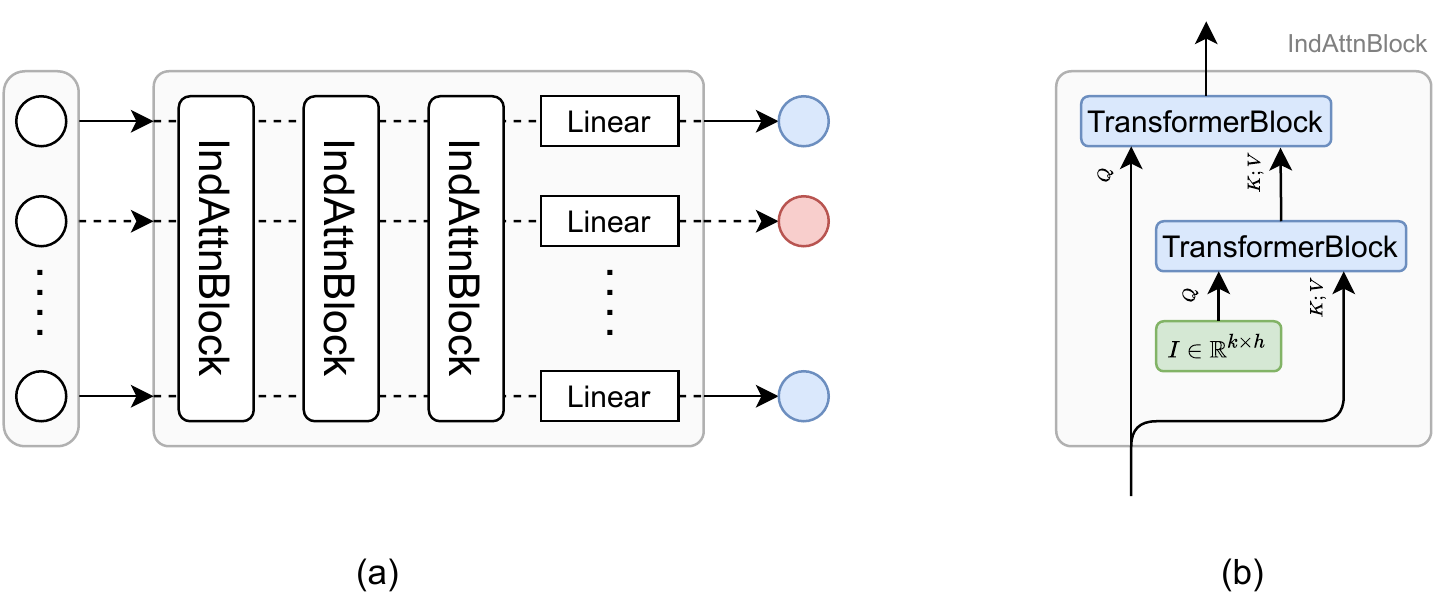}
    \caption{(a) Our network architecture. The input consists of a sample, represented by the event matrix. For every input cell we predict a binary classification label (indicated with colors blue and red). \hi{(b) The induced attention block from eq. \eqref{eq: indattnblock} as introduced in \cite{lee2019set} with the learnable parameters $I$ in green and the TransformerBlock from \eqref{eq: transfblock} in blue.}}
    \label{fig:architecture}
\end{figure}

We start with a brief discussion of the structure of FCM data and then give a detailed description of the network architecture.
\paragraph{FCM Data}
A single sample is represented by a matrix $E \in \mathbb{R}^{N \times m}$ (the {\it event matrix}), where $N$ denotes the number of cells in the sample (typically $10^5 - 10^6$, the exact value for $N$ is different for separate samples) and $m$ denotes the number of markers (typically $10 - 20$ in our case, the exact number depends on the antibodies used). While both the number of cells $N$ and the number of markers $m$ can vary between different samples, there is a set of markers present in every sample (the {\it base panel}). We restrict our method to the markers in this base panel and ignore measurements for other markers, i.e. we keep $m$ fixed and discard measurements of non-base-panel markers. For every index $n \in \{1, \dots, N\}$, $E_n \in \mathbb{R}^m$ is a quantitative representation of the surface markers present on the cell $n$. Ignoring the ordering of cells induced by the FCM machine (i.e. we represent a sample as a set of vectors instead of a sequence) a sample can also be viewed as a bag of features (where a feature is the marker measurement vector for a single cell).
\paragraph{Network Architecture}
\hi{The absence of a low dimensional grid structure (as is typically the case for domains where neural networks excel, like text, where the data is structured on a one-dimensional grid of images that form a two-dimensional grid) makes a direct application of typical neural networks difficult. Self-attention based networks that recently have dominated Natural Language Processing (NLP) related tasks can learn features from sets of embedding vectors (when one ignores the positional embedding that is typically used in NLP problems). However, the memory requirements of such models grow quadratically in the set size \cite{katharopoulos2020transformers} which prevents a direct application to FCM data. To see this, consider the multi-head attention block
\begin{equation}
    \text{Attn}(Q, K, V) = \text{softmax}(Q^T K)V.
\end{equation}
If $Q$ and $K$ derive from the same set of inputs, which is the case for self-attention, the $Q^T K$ multiplication is quadratic in the size set. However, recently an adaption of the self-attention layer has been proposed that reduce the memory requirements from a quadratic growth to linear growth in the set size. In Lee et al. \cite{lee2019set} the standard multi-head self-attention block is replaced by a two-step procedure. For a given input set $X \in \mathbb{R}^{N \times m}$ and $k \in \mathbb{N}$ we initialize a set of parameters $I \in \mathbb{R}^{k \times h}$. Then
\begin{enumerate}
    \item latent features $h \in \mathbb{R}^{k \times h}$ are extracted by performing an attention operation between the set of learnable parameters $I \in \mathbb{R}^{k \times h}$ as query and the input set $X$ as key and value input. 
    \item The resulting hidden features $h$ are used as key and value input for a second attention computation with the input $X$ acting as the query.
\end{enumerate}
We denote this block from now on as Induced Attention Block (IndAttnBlock). This breaks the original $\mathcal{O}(N^2)$ operation into two $\mathcal{O}(N \cdot k)$ operations which circumvents the problem of quadratic complexity (with $k \ll N$ held constant). The latent features are capable of capturing global sample information and the full operation has been proven to be permutation invariant \cite{lee2019set} which justifies the application to set data. We want to point out that, while the network as a whole is permutation invariant, the order of samples in a single forward pass is not mixed up. This is allows us to identify binary classifications in the output with cells in the input. With the multihead attention block from \cite{vaswani2017attention} 
\begin{equation}
    \text{AttnBlock}(X,Y) = \text{LayerNorm}(X + \text{Attn}(X, Y, Y))
\end{equation}
where
\begin{align}\label{eq: transfblock}
    \text{TransfBlock}(X,Y) = \text{LayerNorm}\bigg(&\text{AttnBlock}(X,Y)\\ &+FF(\text{AttnBlock}(X,Y))\bigg)
\end{align}
and the Layernorm from \cite{ba2016layer}, the induced attention block (see Fig. \ref{fig:architecture}, b) can be written as
\begin{equation}\label{eq: indattnblock}
    \text{IndAttnBlock}(X) = \text{TransfBlock}(X, \text{TransfBlock}(I,X)).\footnote{The first Transformer block can be understood as a {\it HopfieldPooling} layer from \cite{ramsauer2020hopfield} while the second block performs the computation of the Layer {\it Hopfield}.}
\end{equation}
Using the induced attention block as a building block, we propose a novel neural network that processes a single sample of FCM data in a single forward pass.
Our network (see Fig. \ref{fig:architecture}, a) is defined as a sequence of three IndAttnBlocks with a row-wise linear layer on top, trained with binary cross-entropy loss.} We do not apply a separate embedding step as in other transformer-based methods but apply our model directly to FCM features (in particular without any positional embedding). We set the number of induced points to $m = 16$, the latent embedding dimension to $d = 32$ and the number of attention heads to $4$ for all three layers.
The resulting model is comparatively lightweight with only $27657$ parameters and can process $\approx 150$ samples/s on an NVIDIA GeForce Titan X\footnote{Only counting the model forward pass, i.e. ignoring time needed for data loading.}.
%
\section{Experiments}
\label{sec: experiments}
We start this section with a brief discussion of the data in subsection \ref{subsec: Data} and the training in subsection \ref{subsec: Training}, followed by the evaluation in subsection \ref{subsec: results}.
\subsection{Data}
\label{subsec: Data}
We evaluate our method on publicly available data\footnote{\href{https://flowrepository.org/id/FR-FCM-ZYVT}{flowrepository.org}} from three different clinical centers. The data consists of bone marrow samples of pediatric patients with B-ALL on day 15 after induction therapy. For all samples ground truth information acquired by manual gating is available for blast and non-blast cells. Table \ref{tab: data} contains an overview of the datasets.
\begin{table}
    \centering
    \caption{Description of the FCM datasets used for experiments.}
    \label{tab: data}
    \begin{tabular}{ cccc }
        \toprule
        {\bf Name} & {\bf City} & {\bf Years} & $\#$\\ [0.5ex] 
        \hline
        {vie14} & Vienna & 2009-2014 & 200 \\ 
        {vie20} & Vienna & 2015-2020 & 319 \\ 
        {vie} & Vienna & 2009-2020 & 519 \\ 
        {bln} & Berlin & 2015 & 72 \\ 
        {bue} & Buenos Aires & 2016-2017 & 65 \\
        \bottomrule
    \end{tabular}
\end{table}
\paragraph{Vienna} The Vienna dataset has been collected at the St. Anna Children’s Cancer Research Institute (CCRI) from 2009 - 2020 with a LSR II flow cytometer (Becton Dickinson, San Jose, CA) and FACSDiva v6.2.  We denote this dataset with {\it vie}, it contains $519$ samples. 
We extract two disjunct datasets from these samples:
\begin{itemize}
    \item {\it vie14:} This dataset contains $200$ samples collected between 2009 - 2014. It is identical to the {\it vie} dataset in \cite{reiter2019automated}. The samples were stained using a conventional seven-colour drop-in panel (“B7”) consisting of the liquid fluorescent reagents: CD20- FITC/ CD10-PE/ CD45-PerCP/ CD34-PE-Cy7/ CD19-APC/ CD38-Alexa-Fluor700 and SYTO 41.
    \item {\it vie20:} This dataset contains $319$ samples collected between 2016 - 2020. The samples were stained using dried format tubes (DuraClone™, "ReALB") consisting of the fluorochrome-conjugated antibodies CD58-FITC/ CD34-ECD/ CD10-PC5.5/ CD19-PC7/ CD38-APC-Alexa700/ CD20-APC-Alexa750/ CD45-Krome Orange plus drop-in SYTO 41.
\end{itemize}
\paragraph{Berlin} The bln Dura \cite{reiter2019automated} (from now on referred to as {\it bln}) dataset contains 72 samples collected in 2016 at Charité Berlin. These samples were recorded with a Navios flow cytometer (Beckmann Coulter, Brea, CA) and assessed by 8-colour multiparameter FCM (“B8”) using a customized dried format tube (DuraClone™, Beckman Coulter) consisting of the seven fluorochrome-conjugated antibodies CD58, FITC/CD10, PE/CD34,\\ PerCPCy5.5/CD19, PC7/CD38, APC/CD20, APC-Alexa750/CD45, Krome-Orange plus drop-in SYTO 41.
\paragraph{Buenos Aires} The bue Dura \cite{reiter2019automated} (from now on referred to as {\it bue}) dataset consists of 65 samples collected between 2016 and 2017 at the Garrahan Hospital in Buenos Aires. The staining panel is identical to the bln Dura set (based on DuraCloneTM cocktail tube; “B8”). The data has been acquired on a FACSCanto II (Becton Dickinson, San Jose, CA) with FACSDiva v8.0.1.
\subsection{Training}
\label{subsec: Training}
We conduct a thorough investigation into cross platform compatibility of our method. For this we train separate models for all four datasets discussed in the subsection above and test the models on every other dataset from Table \ref{tab: data}. The validation datasets for a specific experiment consist of all other datasets (example: for vie14 train and bue test sets we use vie20 and bln as validation data). Additionally we show that our method can be trained on as little as $10$ samples while still reaching competitive results. For these experiments we only use $10$ samples for validation.
For all experiments we use the Adam optimizer \cite{kingma2014adam} with an initial learning rate of $1e-3$ and a Cosine Annealing scheduler \cite{loshchilov2016sgdr} with $10$ iterations and a minimal learning rate of $2e-4$. We train for $100$ epochs with a batch size of $1$ and evaluate on the test with the best checkpoint as measured by the average $F_1$-score on the validation set. We implement our method in Pytorch 1.7.1 \cite{paszke2019pytorch} and use the pre-implemented optimizer and scheduler. Due to the small model size, a training run until convergence on a single NVIDIA GeForce Titan X takes only $\approx 2 - 8$ hours, depending on the dataset size.
\subsection{Results}
\label{subsec: results}
\begin{figure}[ht]
\includegraphics[width=\textwidth]{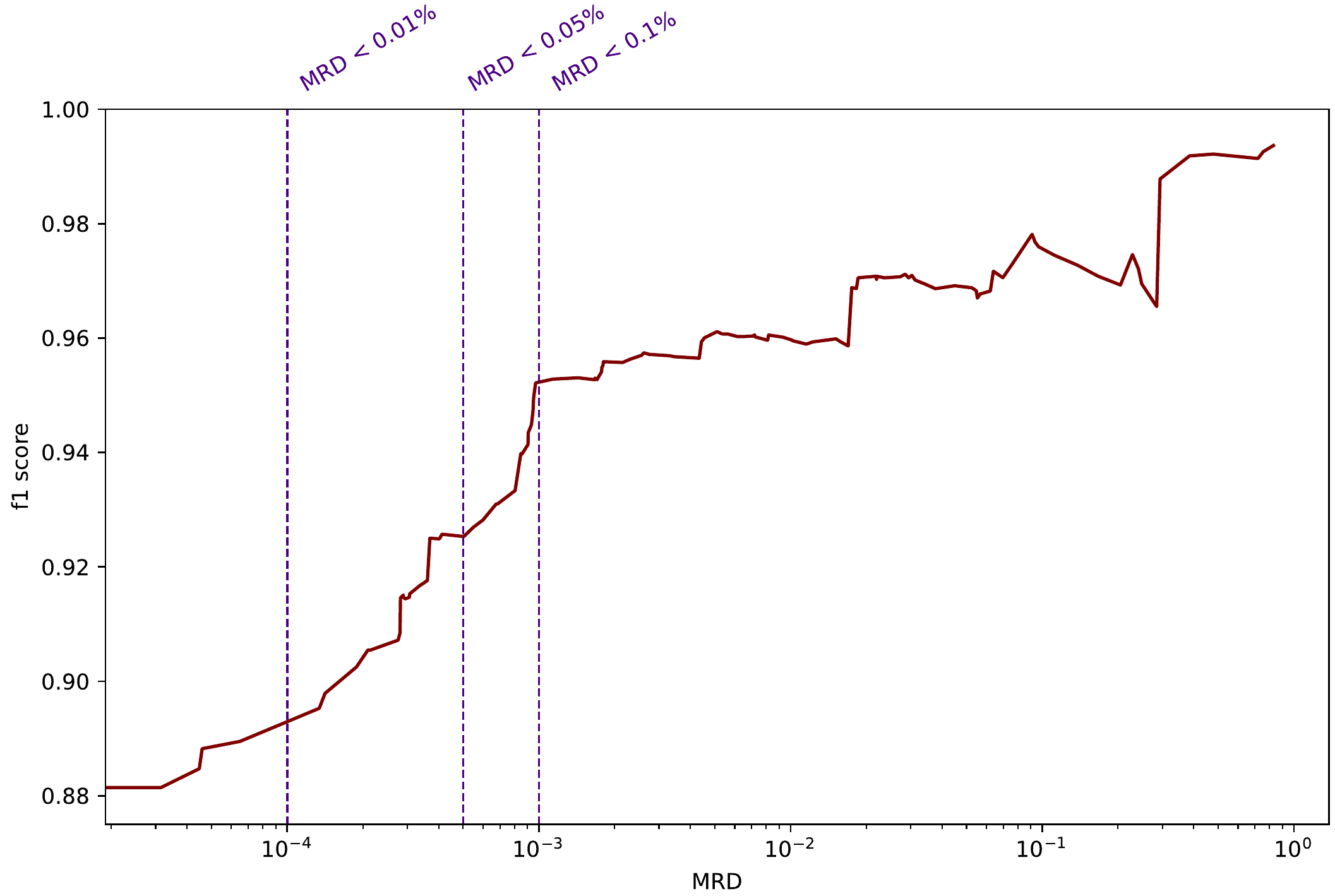}
\caption{Running average of $F_1$ scores against the ground truth MRD values, i.e. for a given MRD value x the line shows the average of $F_1$ scores of all samples within the vie test set with an MRD value larger or equal to x. Due to the logarithmic scale, the $11$ samples in the vie test set with MRD values of $0$ are not shown which explains the missmatch between the lowest running average of $0.88$ and the mean over the whole test set of $0.81$.} \label{fig: mrd f1 plot}
\end{figure}

\begin{figure}[ht]
\includegraphics[width=\textwidth]{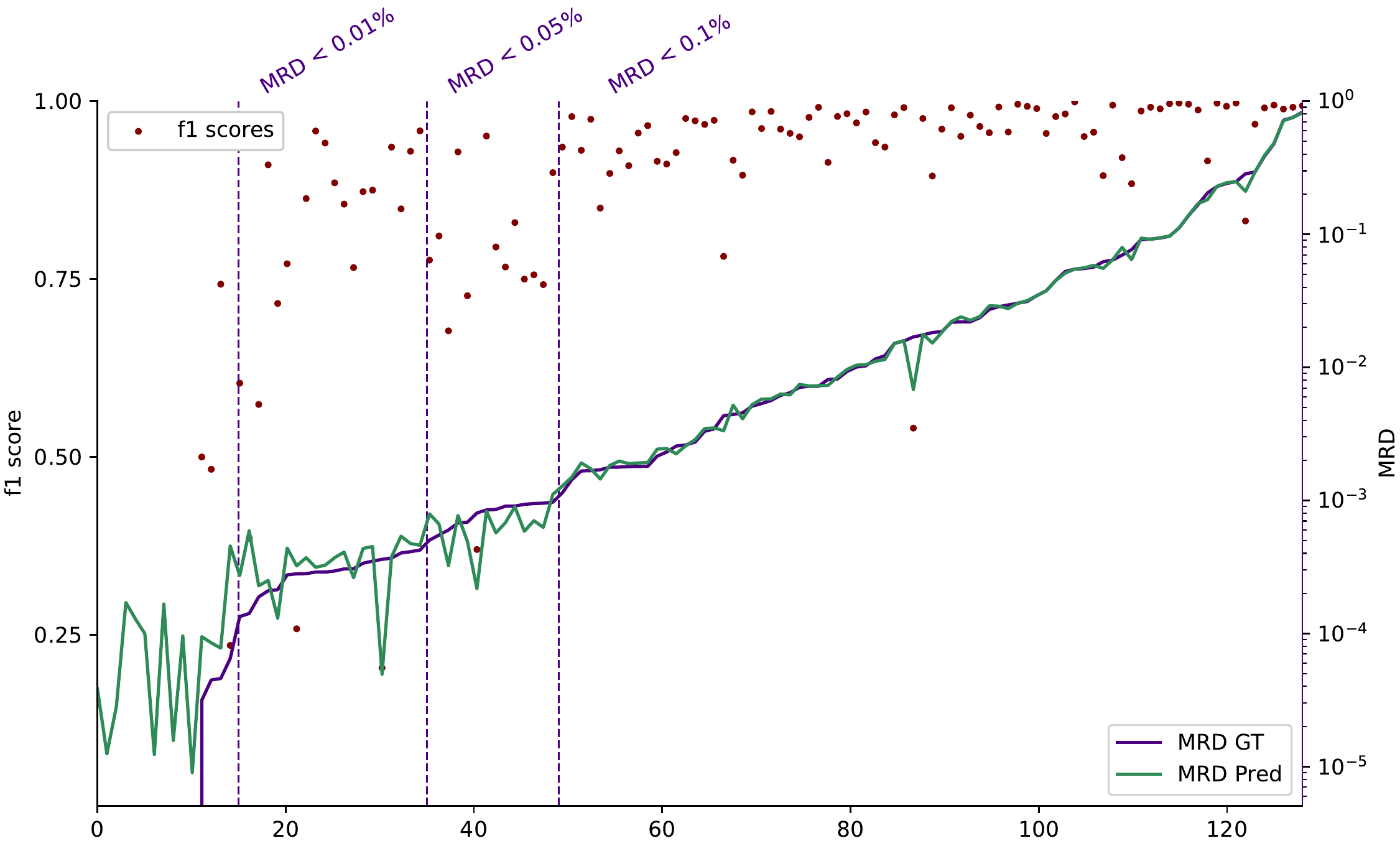}
\caption{$F_1$ scores (red dots), ground truth MRD values (purple lines) and predicted MRD values (green line) for the vie test set.} \label{fig: mrd lineplot}
\end{figure}

\begin{figure}[ht]
\includegraphics[width=\textwidth]{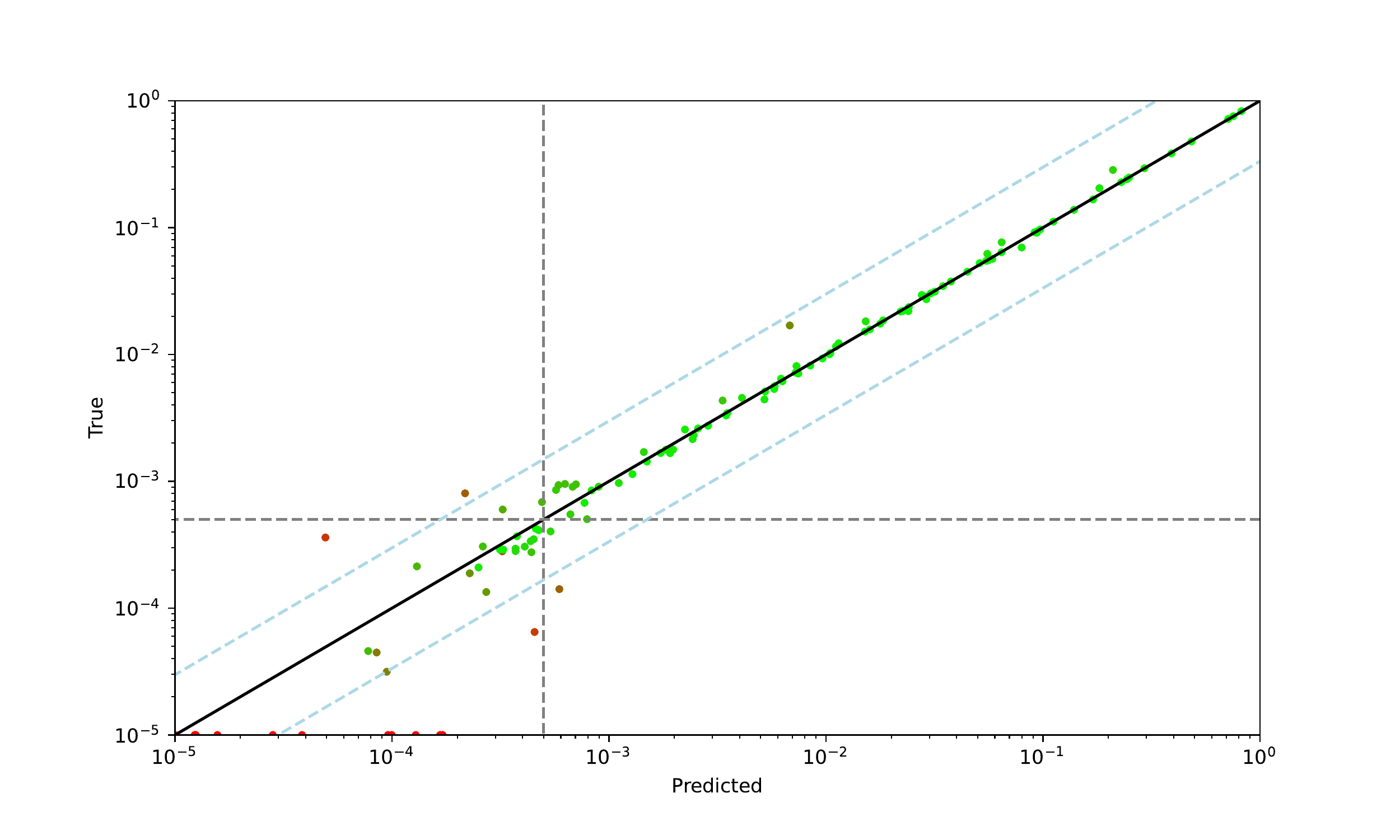}
\caption{$F_1$ scores and predicted MRD values for the test set of the vie experiments. Every dot represents a single sample, with the color donating the $F_1$ score (colors going from red = $0.0$ to green = $1.0$. The dashed lines correspond to MRD values of $5e-4$ which is the lower necessary resolution for patient stratification according to the current international therapy trials of the allied study groups of the iBFM consortium. Predictions that are within the range of either less than $3$ times or more than $1/3$ of the true MRD are considered as acceptable (correct) predictions \cite{dworzak2008standardization}.} 
\label{fig: mrd dotplot}
\end{figure}

For the first set of experiments we train on data from a specific laboratory and test on data from a different laboratory. Exceptions are made for the data from Vienna that we split in 2 sets of data: {\it vie14} (collected between 2009 and 2014) and {\it vie20} (collected between 2015 and 2020). We denote experiments with {\it train/test}, where {\it train} stands for the training set and {\it test} stands for the test set (for example, vie14/bln means we train on vie14 and test on bln). An exception being the {\it vie} experiment where we combine the vie14 and vie20 sets to a single dataset which we randomly split into train, validation and test set. The results of our experiments are listed in Table \ref{tab: results}. To assess the quality of the results we compute average precision (p), average recall (r), average $F_1$ scores (avg $F_1$) and median $F_1$ scores (med $F_1$):
\begin{equation}
    p = \frac{\text{tp}}{\text{tp} + \text{fp}},\; r = \frac{\text{tp}}{\text{tp} + \text{fn}},\; F_1 = 2 \frac{p \cdot r}{p + r}
\end{equation}
where blast cells are "positive" and non-blasts "negative". For samples without blasts or only very few blast cells (see the left-most region in Fig. \ref{fig: mrd lineplot}, In particular the first $10$ samples, where no blasts are present) the $F_1$-score is not a good measure of performance because classification mistakes of single cells can have a significant effect on the $F_1$-score (in particular, for samples with zero blasts, wrongly classifying a cell as a blast cell results in a $F_1$-score of $0$), that is not reflected in clinical significance. We find that because of these reasons, the median $F_1$-score is a better measure for model performance than the average $F_1$-score.
\begin{table}
    \centering
    \caption{\hi{The experimental results evaluated with precision (p), recall (r), average $F_1$-score (avg $F_1$) and median $F_1$-score (med $F_1$). For every experiments we train on all samples from a certain dataset and test on all samples from a different dataset. We compare our method to Reiter et. al. \cite{reiter2019automated}. Boldface values indicate the best performing method for a specific train/test dataset combination.}}
    \label{tab: results}
    \begin{tabular}{ ccccccc }
        \toprule
        {\bf train} & {\bf test} & {\bf p} & {\bf r} & {\bf avg} $\boldsymbol{F_1}$ & {\bf med} $\boldsymbol{F_1}$ & {\bf med} $\boldsymbol{F_1}$~\cite{reiter2019automated}  \\
        \hline
        {vie} & {vie} & 0.81 & 0.83 & 0.81 & 0.94 & - \\
        \vspace{10pt}
        {bln} & {bue} & 0.63 & 0.84 & 0.66 & {\bf 0.87} & {\it 0.68} \\
        {bln} & {vie14} & 0.77 & 0.83 & 0.77 & {\bf 0.90} & {\it 0.35} \\
        {bln} & {vie20} & 0.79 & 0.77 & 0.74 & {\bf 0.87} & {\it 0.48}\\
        \vspace{3pt}
        {bue} & {bln} & 0.56 & 0.92 & 0.62 & {\bf 0.77} & {\it 0.5}\\
        {bue} & {vie14} & 0.76 & 0.88 & 0.79 & {\bf 0.90} & {\it 0.84}\\
        {bue} & {vie20} & 0.79 & 0.74 & 0.72 & {\bf 0.88} & {\it 0.86}\\
        \vspace{3pt}
        {vie14} & {bln} & 0.78 & 0.82 & 0.75 & {\bf 0.9} & {\it 0.81}\\
        {vie14} & {bue} & 0.82 & 0.81 & 0.78 & {\bf 0.95} & {\it 0.84}\\
        {vie14} & {vie20} & 0.81 & 0.74 & 0.73 & {\bf 0.89} & {\it 0.86}\\
        \vspace{3pt}
        {vie20} & {bln} & 0.64 & 0.87 & 0.66 & {\bf 0.81} & {\it 0.25}\\
        {vie20} & {bue} & 0.82 & 0.69 & 0.71 & {\bf 0.86} & {\it 0.81}\\
        {vie20} & {vie14} & 0.82 & 0.69 & 0.71 & 0.86 & {\it {\bf 0.89}}\\
        \bottomrule
    \end{tabular}
\end{table}
\hi{We compare our method to the GMM based model described in Reiter et al. \cite{reiter2019automated} which we evaluated on the vie14, vie20, bln and bue datasets. The complete set of results for the conducted experiments can be seen in Table \ref{tab: results}. We outperform the existing approach \cite{reiter2019automated} in all train/test combinations except for vie20/vie14 where we reach comparable results. Our method is significantly faster with inference times of ~$5$ms for our method vs ~$3000$ms for the GMM based approach \cite{reiter2019automated}. We consistently reach median $F_1$ scores $\geq 0.86$ with the only exceptions being bue/bln with $0.77$ and vie20/bln with $0.81$.}
For bue/bln in particular our method only reaches a median $F_1$ score of $0.77$ with a precision of $0.56$ and recall of $0.92$ indicating that for sufficiently different data sources the performance can degrade. However, adding $5$ random samples from the test set to the training set and testing on the remaining samples improves median $F_1$ score, precision and recall to $0.87$, $0.7$ and $0.91$ indicating that if a small number of labeled data is available cross-laboratory performance of our method can be improved significantly. In general we find that when measured with respect to the $F_1$ score, our method performs better for samples with larger MRD values. Figure \ref{fig: mrd f1 plot} shows the average $F_1$ score for all samples with an MRD value above the threshold given by the value on the x axis for the vie test set. Samples with low $F_1$ score are predominantly those with a smaller MRD i.e. lower count of blast cells. For low MRD values our method tends to overestimate the true value more often than it underestimates it. This can be seen in Figure \ref{fig: mrd dotplot} where the ground truth MRD is plotted against the predicted MRD. A different visualization is given by Figure \ref{fig: mrd lineplot} where for every sample the true MRD, the predicted MRD and the $F_1$ score are given. 
%
\section{Conclusion}
\label{sec: conclusion}
In this work, we proposed a novel method for automated identification of cell populations and used it for the detection of blast cells in B-ALL FCM data. Our method is based on a lightweight ($27657$ parameters) neural network that allows fast ($\approx 150$ samples/s) processing of samples with $10^5 - 10^6$ cells on a NVIDIA GeForce GTX TITAN X. We trained the model in a supervised manner on as few as $65$ samples of data from three different sources and showed that our method is capable of generalizing to unseen data. Our method is different from existing approaches that utilize neural networks for automated FCM analysis \cite{scheithe2019monitoring,licandro2018wgan} in that we make use of self-attention layers that allow the network to attend to all cells in the sample at once instead of processing every cell independently. For future work, we argue that data augmentation methods that capture device differences (for example as proposed in \cite{shaham2017removal} for mass spectroscopy) would help improve generalization (like the performance drop for the vie20/bln experiment).
%
\section{Acknowledgement}
\label{sec: acknowledgement}
We thank Dieter Printz (FACS Core Unit, CCRI) for flow-cytometer maintenance and quality control, as well as Daniela Scharner and Susanne Suhendra-Chen (CCRI), Jana Hofmann (Charité), Mariann eDunken (HELIOS Klinikum), Marianela Sanz, Andrea Bernasconi, and Raquel Mitchell (Hospital Garrahan) for excellent technical assistance. We are indebted to Melanie Gau, Roxane Licandro, Florian Kleber, Paolo Rota and Guohui Qiao (all from TU Vienna) for valuable contributions to the AutoFLOW project. We thank Markus Kaymer and Michael Kapinsky (both from Beckman Coulter Inc.) for kindly assisting in the provision of customized DuraCloneTm tubes for this study as designed by the authors. Notably, Beckman Coulter Inc. did not have any influence on study design, data acquisition and interpretation, or manuscript writing. The study has received funding from the European Union’s H2020 Research and Innovation Program through Grant number 825749 “CLOSER: Childhood Leukemia: Overcoming Distance between South America and Europe Regions”, the Vienna Business Agency under grant agreement No 2841342 (Project MyeFlow) and by the Marie Curie Industry Academia Partnership \& Pathways (FP7-MarieCurie-PEOPLE-2013-IAPP) under grant no. 610872 to project “AutoFLOW” to MND. The authors acknowledge TU Wien Bibliothek for financial support through its Open Access Funding Programme.

\section{Declarations of interest}
Michael N. Dworzak received payments for travel, accommodation or other expenses from Beckman-Coulter. The other authors declare no competing financial interests.
\newpage
%
%
%
\bibliographystyle{elsarticle-num} 
\bibliography{references}
\end{document}